# Room-temperature generation of heralded single photons on silicon chip with switchable orbital angular momentum

SHAN ZHANG[1], XUE FENG[1,*], WEI ZHANG[1], KAIYU CUI[1], FANG LIU[1], AND YIDONG HUANG[1]

[1] *Department of Electronic Engineering, Tsinghua University, Beijing, China*
* *x-feng@tsinghua.edu.cn*

**Abstract**

In quantum optics, orbital angular momentum (OAM) is very promising to achieve high-dimensional quantum states due to the nature of infinite and discrete eigenvalue, which is quantized by the topological charge of *l*. Here, a heralded single-photon source with switchable OAM modes is proposed and demonstrated on silicon chip. At room-temperature, the heralded single photons with 11 OAM modes ($l$=2~6, -6~-1) have been successfully generated and switched through thermo-optical effect. We believe that such an integrated quantum source with multiple OAM modes and operating at room-temperature would provide a practical platform for high-dimensional quantum information processing. Moreover, our proposed architecture can also be extended to other material systems to further improve the performance of OAM quantum source.

## 1. Introduction

The orbital angular momentum (OAM) is an independent degree of freedoms (DoFs) of photons, which corresponds to the spiral phase fronts, with the quantized eigenvalue as the topological charge (*l*)[1]. Benefiting from the unique characteristics of the infinite and discrete topological charges, OAM modes are natural for generating high-dimensional states and could increase the information capacity carried by photons[2–5]. Especially in quantum domain, it is strongly desired to increase the information content encoded on a single photon for higher efficiency, better noise resistance, richer resources and more flexibility of quantum communications, computations and simulations[6–9]. So far, OAM encoded quantum states have been utilized in quantum key distribution[10–13], quantum ghost imaging[14,15] and high-dimensional quantum entanglement[16,17]. The potential application of high-dimensional quantum information have stimulated the research on generating[18,19], manipulating[20,21] and detecting[22,23] the optical quantum states encoded with OAM modes.

As an effective way to achieve higher dimensional quantum states, OAM single-photon sources are mainly demonstrated on the spontaneous parametric down-conversion (SPDC) in nonlinear crystals[24–27], where bulky spatial optical systems are required. Clearly, an integrated OAM quantum source is preferred for its compactness, high stability and flexibility. The pioneer work of the integrated OAM single-photon source is demonstrated with InAs/GaAs epitaxial quantum dots embedded in a micro-ring cavity, in which the OAM modes with |*l*|=6 have been successfully generated at 30K[28]. To utilize the high-dimensional feature, the usable OAM modes should be as more as possible after considering both applications of multiplexing and dynamic coding [17], and the latter also requires that the topological charges could be dynamically switched. We have recently demonstrated a heralded single-photon source with switchable OAM modes, in which 5 OAM modes ($l$=3~7) can be obtained with fixed incident wavelength[29]. However, such work is based on the spontaneous four-wave mixing (SFWM) in dispersion-shifted fiber (DSF) so that it is not a fully on-chip device and cooling to 77K is still required to reduce the Raman scattering noise in DSF[30]. Till now, an integrated quantum light source with switchable OAM modes and operating at room-temperature has not yet been realized.

To tackle the aforementioned two challenges, we have proposed and demonstrated an integrated heralded single-photon source with 11 switchable OAM modes operating at room-temperature in this work. The whole device is integrated on a monolithic silicon chip and spatially divided to two parts: generating correlated photon pairs and then transforming to OAM modes. Here, the heralded single-photon source is based on the SFWM in a silicon wire waveguide (SW-WG) as a result of its high nonlinearity and narrow Raman bandwidth[31–33]. The noises caused by Raman scattering could be filtered without the cryogenic requirement so that the whole device could operate at room-temperature. For mode transformation, an integrated OAM emitter based on the shallow ridge waveguide (SR-WG) could load the OAM modes onto single photons[34]. Benefitting from the spatial separation of the source and emitter, the room-temperature operating condition, and the high thermo-optical coefficient of silicon, the topological charges of generated OAM modes can be switched through the thermo-optical effect[35] while the performance of single-photon emission is not disturbed. Assembling both the heralded single-photon source and OAM emitter by a well-designed transition structure, the heralded single photons with switchable OAM modes ($l$=2~6, -6~-1) are obtained. The coincidence counts (CCs) and coincidence counts-to-accidental coincidence counts ratio (CAR) are measured as 27.5~82 and 16.33~57.63 within 10 minutes for 11 OAM modes, respectively. It should be mentioned that there is no obvious deterioration of the CAR value compared with that of SW-WG alone (34.09~52.05). Thus, the correlation between photon pairs is kept after the mode conversion of OAM emitter. We believe that this work is essential for exploring the application of OAM modes in high-dimensional quantum information processing.

## 2. Results
## 2.1 Principle

The whole device is fabricated on the silicon platform for both high nonlinearity and thermo-optical coefficient. The silicon on insulator (SOI) substrate with top silicon thickness of 220nm is utilized here. As shown in Fig.1, to generate the single-photon state with switchable OAM modes on chip, our proposal is to utilize the spatially separated single-photon source and OAM emitter. With proper structural design, a heralded single-photon source could be realized through degenerate SFWM in SW-WG (yellow section in Fig.1), where correlated photon pairs are generated at room-temperature[31,32,36]. Each photon pair can be divided into signal and idler photon according to their wavelengths. Through the wavelength-selecting feature of ring cavity in the OAM emitter, the signal photon would be coupled into the emitter and converted to OAM modes, while the idler photon is scattered from the output grating directly. Thanks to the room-temperature operation of heralded single-photon source as well as the spatial separation of the source and emitter, a thermo-optical controller attached to OAM emitter can be utilized to switch the topological charges of generated OAM modes. Finally, the CCs and CAR between signal and idler photon could confirm the successful generation of the heralded single photons with switchable OAM modes. Fig. 2(a) is the microphotograph of the fabricated device and the pump laser is coupled into SW-WG through the vertical input grating (shown as the inset in left bottom of Fig. 2(a))

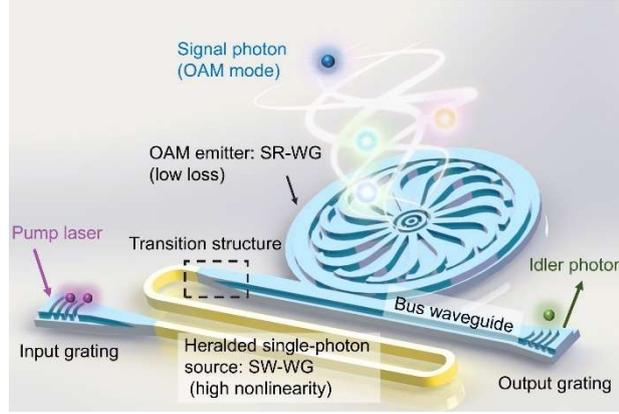

**Fig.1 The schematic of integrated heralded single-photon source with switchable OAM modes.** SW-WG: silicon wire waveguide; SR-WG: shallow ridge waveguide. Such two types of waveguides are utilized due to the features of the low transmission loss and high nonlinearity, respectively. The device is based on the silicon platform and the different colors are used to distinguish different waveguides (SW-WG and SR-WG).

For most quantum light sources based on nonlinear processes, Raman scattering noise is the main factor deteriorating the signal-to-noise ratio (SNR). It is worth mentioning that, due to the crystal nature of silicon material, the linewidth of Raman scattering is as narrow as 103GHz and located 15.6THz away from the pump frequency[31]. Thus, the Raman noise could be readily filtered out at room-temperature. In order to obtain a higher nonlinear coefficient, the width and height of the SW-WG is designed as 460nm×220nm[32] whose cross profile is shown in the Fig.2 (b). Moreover, the length of SW-WG is considered as the trade-off between the correlated photon-pair generation rate and transmission loss. After some experimental measurements, the length is set as 5mm (see Fig.S1 in supplementary information). For generated photon pairs, we take the signal photon as heralded single photon and idler photon as heralding single photon in this work. Actually, the heralded single photon is post-selected by CCs between these two photons. Meanwhile, the calculated CAR can evaluate the SNR of the heralded single-photon source[37–39].

The design of OAM emitter is based on our previous work[34]. It consists of a bus waveguide, micro-ring cavity, 32 azimuthally uniformly distributed download waveguides and scattering grating in the center of the cavity. The download waveguides would pick up part of the whispering gallery mode (WGM) around the micro-ring cavity, and collect them to the scattering grating (Fig.2 (d)). Conditioned on the intrinsic angular momentum of WGM, the scattered mode would carry the OAM. When the number of download waveguides is fixed, the topological charges of generated OAM modes directly depend on the order of WGM (details are shown in supplementary information). Thus, we can vary the order of WGM with thermo-optical controller (titanium electrode above micro-ring cavity), in turn switching the topological charges at fixed incident wavelength. To improve the efficiency of thermo-optical controller, the radius of micro-ring cavity is designed as 200μm for a narrower free spectral range (FSR) to ~0.5nm. As shown in the Fig. 2 (c), for a lower transmission loss, the SR-WG is utilized for the OAM emitter. The etching depth and width of SR-WG are 70nm and 1μm, respectively. Obviously, there are two types of waveguides corresponding to the heralded single-photon source and OAM emitter, respectively. Hence, a transition structure is designed between them so that the SW-WG is gradually transformed to SR-WG within a length of 10μm[40]. Fig.2 (e) shows the gradually enlarged SR-WG in transition structure. It should be mentioned that the thermo-optical controller is also important for precisely tuning the operating wavelength of the OAM emitter according to the wavelength of signal photons. By properly setting the wavelengths of the pump and signal/idler photons, together with the thermo-optical controller, the signal photons can be coupled into the OAM emitter and converted to OAM modes while the pump and idler photons are not.

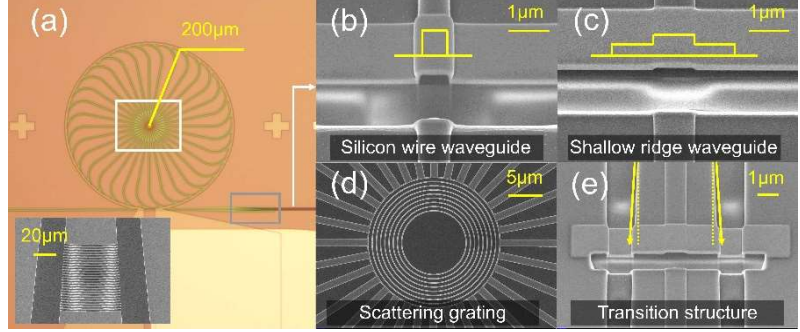

**Fig.2 Photographs of the fabricated device. a.** The microphotograph of the whole device and the inset is the scanning electron microscope (SEM) photograph of the input grating (same as output grating); **b.** The SEM photograph of silicon wire waveguide, **c.** shallow ridge waveguide, **d.** scattering grating (marked by white box in **a**) and **e.** transition structure (marked by gray box in **a**). The period and duty cycle of scattering grating are 630nm and 50%, same as the input (output) grating.

## 2.2 The characteristics of OAM emitter

To determine the arrangement of pump, signal and idler wavelengths, the transmission spectrum of the OAM emitter is measured at the beginning. A tunable laser is coupled into the emitter through the input grating while the laser power is set as 0dBm to alleviate the nonlinear loss in SW-WG. Meanwhile, the driving voltage of thermo-optical controller is set as 0V and the intrinsic transmission spectrum of the emitter could be obtained by scanning the resonant wavelength of the OAM emitter. As shown in Fig.3 (a), the free spectrum range (FSR) and full-width half-maximum (FWHM) are ~0.5nm and ~0.045nm, respectively. The corresponding relation between the resonant wavelength and topological charge could be identified through the interference patterns of the generated OAM modes and Gaussian mode[41] (see Fig.S2 in supplementary information). There are some typical topological charges are marked below the corresponding resonant wavelength. To avoid the nonlinear process in micro-ring cavity, which would introduce unexpected noises, the pump photons should directly propagate through the bus waveguide and be scattered from the output grating as well as the idler photons. Thus, after considering both resonant wavelengths and the group velocity dispersion of SW-WG, the wavelength of pump laser ($\lambda_p$) is set as 1552.5nm. Moreover, the signal wavelength has to be aligned with the resonant wavelength of micro-ring cavity while the idler wavelength does not and both of them should follow the relation for SFWM as $\omega_s + \omega_i = 2\omega_p$ ($\omega_s$, $\omega_i$ and $\omega_p$ are angular frequencies of signal, idler and pump photons). As shown in the Fig.3, for OAM mode with $l=-6 \sim -1$, the signal wavelength ($\lambda_s$) is set as 1547.72nm (marked as green dashed line in Fig.3) while correspondingly, the idler wavelength ($\lambda_i$) is 1557.36nm (marked as blue line in Fig.3). For OAM modes of $l=2\sim 6$, $\lambda_s$ and $\lambda_i$ are exchanged, since the resonant wavelength of $l=2$ is closer to 1557.36nm (more discussion are shown in supplementary information). Moreover, the spectrum would be red-shifted as increasing the driving voltage so that an arbitrary resonant wavelength could always be aligned with the fixed signal wavelength. As an example, the spectrum as driving voltage equals 13.97V is shown in Fig.3 (b), in which the OAM mode of $l=4$ is aligned with $\lambda_s$ while correspondingly, $\lambda_i$ is misaligned with the resonant wavelength.

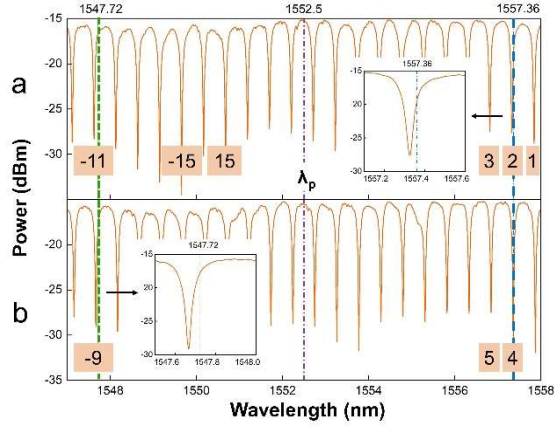

**Fig.3 The transmission spectrum of the OAM emitter with driving voltage of a. 0V and b. 13.97V.** The insets are the zoom-in figures for more clarity. Some typical topological charges corresponding to the resonant wavelength near $\lambda_s$ and $\lambda_i$ are marked in orange boxes.

The emission efficiency and mode purity of OAM emitter are also evaluated with the tunable laser. Here, the incident wavelength is same as $\lambda_s$. Specifically, OAM modes with $l$=-6~-1 and $l$=2~6 are measured at the incident wavelength of 1547.72nm and 1557.32nm, respectively. With a fixed incident wavelength, the topological charges could be switched by varying the driving voltage on thermo-optical controller. The emission efficiency is defined as $\frac{P_{out}}{P_{in}}$. Here, $P_{in}$ is the optical power coupled into the OAM emitter, which can be calculated as the input power subtracting the vertical coupling loss and the insertion loss of the SW-WG. $P_{out}$ is the optical power collected directly above the emitter. According to our experiments, the losses of vertical coupling and SW-WG are estimated as 7dB and 1dB, respectively. Then, the emission efficiencies versus varied topological charges are shown as the blue stars in Fig.4. Moreover, the mode purity of the generated OAM mode with topological charge of $l$ is defined as $\frac{P_l}{\sum_{m=-6}^{6} P_m}$, in which $P_x$ ($x=l/m$) is the power of the OAM mode with topological charge of $l$ or $m$. Here, the topological charge range of $m$=-6 ~ 6 is considered as the bases. It should be mentioned that the generated OAM modes are azimuthally polarized which can be seen in the insets of Fig. 4[42]. The doughnut intensity profile in first column corresponds to the OAM mode with $l$=0 and the other four images are the intensity cross-sections after passing through a polarizer with different directions as marked by the yellow arrows. The azimuthal polarization could be equally decomposed to left-handed and right-handed circular polarized (LHCP and RHCP) components with the topological charge donated as $l_L$ and $l_R$. They satisfy the relation of $l=l_L-1=l_R+1$[34]. Here, the values of mode purity have been averaged with the measured values of both LHCP and RHCP components and shown as yellow dots in the Fig.4 (see Fig.S5 in supplementary information).

Theoretically, a higher mode purity could be guaranteed if the power extracted by each downloaded waveguide is as equal as possible. Hence, there is a trade-off between the emission efficiency and mode purity. In this work, the mode purity is taken more concern than the emission efficiency so that the coupling efficiency of each downloaded waveguide is set as low as 1%. As shown Fig.4, most of the measured mode purities are larger than 80%. Correspondingly, the values of emission efficiency are ranging between 0.93% ~1.97%. It should be noticed that the emission efficiencies would decrease with the increasing absolute values of topological charge. According to the Nyquist theory, the spatial sampling rate would be low for OAM modes with high $|l|$ due to the fixed number of downloaded waveguides, which would introduce a lower emission efficiency. The small fluctuations in measured values may be related to the deviation of chip fabrication.

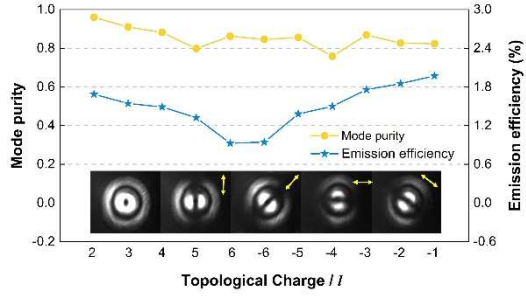

**Fig.4. The mode purities and emission efficiencies of 11 OAM modes.** The insets are the intensity cross-sections of the generated OAM mode with *l*=0 before (the first one) and after a polarizer. The directions of the polarization for the latter four insets are marked by yellow arrows.

### 2.3 The coincidence counts of silicon wire waveguide

At first, the performance of the SW-WG serving as the heralded single-photon source is evaluated. As shown in the Fig.3 (a), without driving voltage, all of the pump, signal and idler photons are not coupled into the OAM emitter so that the CCs of the SW-WG could be measured from the output grating. Here, the pump is a pulsed laser (40MHz, 1552.5nm), coupled into the SW-WG from the input grating, and then the correlated photon pairs are generated through SFWM process. Afterwards, such photon pairs would directly propagate through the bus waveguide and be scattered out of the chip from the output grating. With the help of vertical coupling, the photon pairs are coupled into a fiber-based dense wavelength division multiplexer (DWDM). Thus, the signal and idler photons are separately guided into two single photon avalanche diodes (SPADs) connected by time-correlated single photon counting (TCSPC) for CCs measurement. The typical CCs of SW-WG within 10 minutes is shown in the Fig.5 (a), there is a main peak of CCs and some equidistantly distributed small side peaks, which can be clearly observed in the enlarged figures as Fig.5 (b) and (c), respectively. In Fig.5 (b), the main peak is actually broadened due to the time jitter of SPADs[43]. Thus, the CCs are calculated as the sum of the five peaks within a time slot of 320ps (the red dashed box in Fig.5 (b)) and the result is 77.9K counts/10mins. Different from the background noises caused by the dark counts of SPADs, the time interval of the small side peaks corresponds to the period of pulsed laser (25ns) as shown in Fig.5 (c)[44]. For a heralded single-photon source, CAR is defined as the ratio of CCs to accidental coincidence counts (ACCs). Here, the ACCs are estimated as the value of 14 side peaks evenly distributed around the main peak while the time jitter is also considered. As a result, 14 values of CAR are obtained and the maximum and minimum are 52.05 and 34.09, respectively.

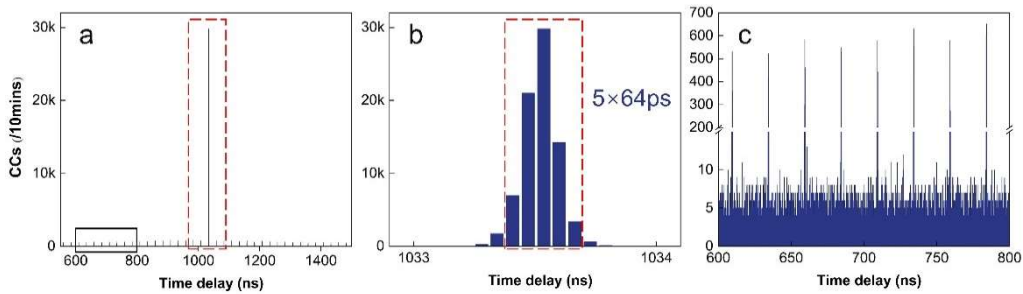

**Fig.5. The coincidence counts (CCs) of silicon wire waveguide. a.** A typical result of CCs. **b.** The zoomed-in view of main peak corresponding to the red dashed box in (a). **c.** The zoomed-in view of side peaks with the time delay of 600~800ns corresponding to the black box in (a).

### 2.4 The coincidence counts of heralded single photons with switchable OAM modes

With a proper driving voltage, the signal photons could be coupled into the OAM emitter and converted to OAM modes. When adjusting the driving voltage, the topological charge of OAM mode can be switched with the $\lambda_s$ fixed at 1557.36nm for *l*=2~6 and 1547.72nm for *l*=-6~-1, respectively. Shown as the orange stars in Fig.6 (a) and (b), there is a linear relation between the topological charge and driving power. Actually, the resonant order of WGM would increase with higher driving power so that the corresponding topological charge could be switched. Specifically, shown as the green squares in Fig.6 (a) and (b), for $\lambda_s$= 1557.36nm, *l*=2~6 is corresponding to the resonant order of 282~286, while *l*=-6~-1 is corresponding to 306~311 for $\lambda_s$= 1547.72nm.

To verify the generation of heralded single photons with switchable OAM modes, CCs and CAR corresponding to 11 OAM modes are measured within 10 minutes. As shown in the Fig.6 (c) and (d), CCs of 27.5~82 and CAR of 16.33~57.63 are obtained for different OAM modes. Here, the time slot and calculation method of CAR are same as the conditions of SW-WG alone except that, for OAM modes, the final CCs and CAR are the average of those of both RHCP and LHCP components. For CCs shown in Fig. 6 (c), there is an obvious attenuation compared with SW-WG alone (CCs of 27.5~82 versus ~77.9k within 10 minutes). This mainly comes from the insertion loss of OAM emitter and detection setup, which would both increase as the |$l$| rising (see Fig.S6 in supplementary information). Hence, the CCs shown in Fig. 6(c) goes down for $l$=2~6 and then up for $l$=-6~-1. For CAR shown in Fig. 6 (d), two orange horizontal lines represent the range for SW-WG alone and most CARs of single photons with OAM states fall in this range. This indicates that the correlation between photon pairs has been kept after the modulation of OAM emitter. Moreover, these results demonstrate that the heralded single-photon source with switchable OAM modes has been successfully realized at room-temperature. Benefit from the integrated scheme and crystal nature of silicon, the overall CCs and CAR have been improved significantly compared to our previous work[29].

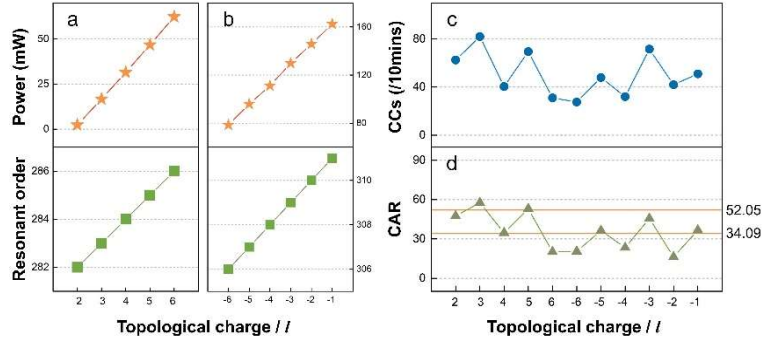

**Fig.6. The characteristics of heralded single-photon source with switchable OAM modes. a.** and **b.** The driving power (orange stars) and the corresponding resonant order (green squares) versus topological charge ranging within $l$=2~6 and $l$=-6~-1, respectively; **c.** The CCs of the heralded single photons with switchable OAM modes; **d.** The CAR of single photons with OAM compared to that of SW-WG alone (marked as the two orange lines with values of 34.09 and 52.05, respectively).

In our experiments, the CCs of heralded single photons with OAM modes are measured through the SPADs packaged by single mode fiber (SMF). Thus, a SLM is utilized to convert the quantum OAM modes to fundamental Gaussian mode. Since the SLM requires linear incident polarization, a quarter wave plate (QWP) and a polarizer are set in detection system to convert the polarization of generated OAM modes. Together with all this devices above and the objective lens for collection (coupling efficiency 40%), the losses in detection process are estimated to be -14.53~-18.96dB for different OAM modes (see Fig.S6 in supplementary information). Actually, such additional loss introduced by detection can be avoided in practical applications. It means that the usable CCs of heralded single photons with OAM modes could be estimated as the value before detection setup, which is as high as 1.2k~3.2k/10mins. The details of the CC values corresponding to 11 OAM modes are shown in Fig 7. It can also be seen that the variation trend of the estimated CCs is very similar to that of the emission efficiency shown in Fig.4.

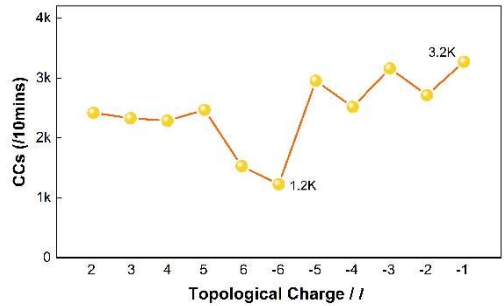

**Fig.7. The estimated coincidence counts (CCs) of heralded single photons after deducting the insertion loss of detection system.** The maximum and minimum values are 3.2k and 1.2k corresponding to the l=-1 and -6, respectively.

## 3. Discussion

In this work, the whole device consists of two main units: the heralded single-photon source based on SW-WG and the integrated OAM emitter based on SR-WG. Essentially, this is a kind of heterogeneous structure assembled on a single chip to achieve both of the high nonlinear coefficient and thermo-optical coefficient. Such two characteristics are exactly beneficial to efficiently generating heralded single photons and widely switching topological charges. Actually, the same structural design could be utilized on other material platform. Particularly, the CCs can be improved by utilizing materials with higher nonlinearity and faster switching speed can be achieved with other physical mechanism. For example, based on the lithium niobate platform, both high nonlinear coefficient[45] and high electro-optical coefficient[46] can be obtained to realize higher CCs and faster switching of topological charges. Furthermore, the spatial separation of the source and emitter provides more freedom for structural design. These two units could also be achieved based on two different materials on the same chip. The primary consideration is that the high nonlinear coefficient is essential to heralded single-photon source while low loss and feasible tunability are for OAM emitter.

## 4. Method

**Fabrication of the device.** There are two rounds of etching corresponding to the SW-WG and SR-WG (of OAM emitter) fabricated through deep (220nm) and shallow (70nm) etching, respectively. Then, the titanium and aluminum are successively deposited on the top of the OAM emitter as the heat resistor and conductive electrode. The whole device is attached to a printed circuit board (PCB) for voltage supply.

**CCs and CAR.** The experimental setup is demonstrated in the Fig.8. After modulated to required wavelength (1552.5nm) and polarization, the pulsed laser is coupled into the device through vertical coupling grating. When applying appropriate driving voltage, the idler photons would be scattered from output grating and detected by SPAD directly. The signal photons with OAM modes are collected by objective lens and modulated by wave plates and SLM. Considering the polarization dependence of SLM, the azimuthal polarization of OAM modes would be converted to linear polarization by a QWP and a polarizer. After converted to fundamental Gaussian mode by the phase mask on SLM, the signal photons are directed into SPAD as well. Finally, the CCs between idler and signal photons could be obtained through TCSPC. As tuning the driving voltage, the heralded single photons with different OAM modes could be generated and detected.

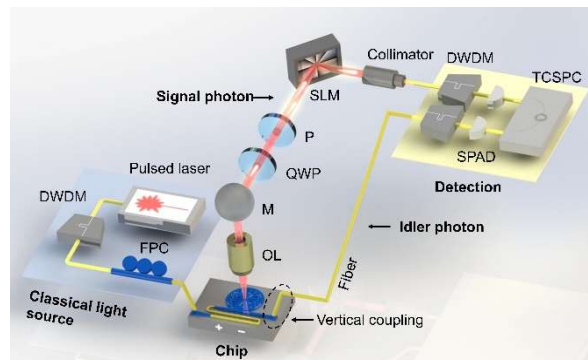

**Fig.8. The schematic for measuring the CCs and CAR of heralded single-photon source with OAM modes.** DWDM, Dense wavelength Division Multiplexer; FPC, fiber polarization controller; OL, objective lens; M, mirror; P, polarizer; SPAD, single photon avalanche diode; TCSPC, time-correlated single photon counting.

# Supplementary information

## 1. Optimizing structural parameters of silicon wire waveguide

In this work, the dimensions of silicon wire waveguide (SW-WG) are selected as 460nm×220nm×5mm (width×height×length). Firstly, the height is fixed at 220nm according to thickness of top silicon of the SOI substrate. Then, the width is determined by the single-mode condition and as small as possible mode profile to obtain a higher nonlinear coefficient[1]. Considering the fabrication deviation, the width of SW-WG is set as 460nm ultimately. For the length, it would be a trade-off between high nonlinearity and low transition loss. To determine the value, we have fabricated three lengths of SW-WGs: 3mm, 5mm and 8mm. As shown in Fig. S1, the coincidence counts (CCs) of such three SW-WGs are measured in one minute. According the CCs, the SW-WG with length of 5mm is adopted to achieve relatively high CCs. Here, the pump/signal/idler wavelengths are 1552.5nm, 1557.36nm and 1547.74nm and the pump source is a pulsed laser (40MHz, the average power equals 0.4mW).

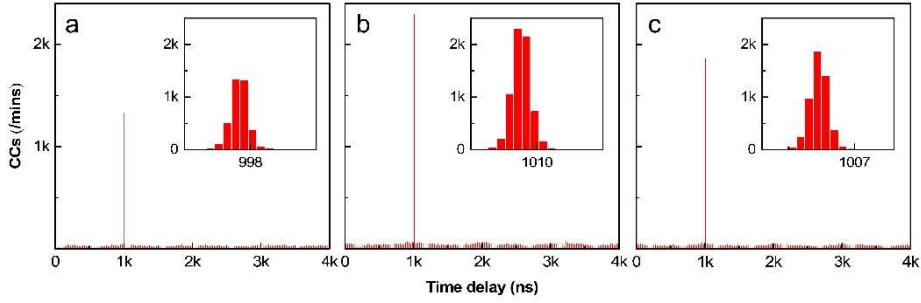

Fig. S1. The coincidence counts of silicon wire waveguides with three lengths in one minute. a: 3mm; b: 5mm; c. 8mm. The insets are the zoomed-in view of main peaks.

## 2. The operation principle of OAM emitter

The generated OAM modes originate from the whispering gallery mode (WGM) in micro-ring cavity. There are 32 download waveguides (DWs) azimuthally distributed in this cavity, extracting the power from WGM and converging to the centric scattering grating. Since all DWs are equally spaced and the coupling distances with cavity are same (350nm), they could pick up part power of the WGM with uniformly phase gradient and nearly equal strength. The phase difference between adjacent waveguides is defined as[2]:

$$\Delta\phi = \frac{2\pi N}{M} \mod 2\pi - \pi \in [-\pi, \pi),$$

where N is the order of WGM and M=32 is the number of DWs. Furthermore, the topological charge ($l$) can be derived from:

$$l = \frac{\phi}{2\pi} = \frac{M\Delta\phi}{2\pi} \in \mathbb{Z}, \Delta\phi \neq -\pi$$

Hence, when M is fixed, the topological charge of generated OAM modes are directly depends on the order of WGM. According to the equation of WGM, the order could be switched by varying the effective refractive index of micro-ring cavity as incident wavelength is fixed. In this work, the thermo-optical effect is utilized to achieve this function.

## 3. Identify the topological charge through interference patterns

Light beams carrying OAM have spiral phase structures, whose handedness and magnitude are described by the topological charge $l$[3]. The interference patterns between OAM modes and the fundamental Gaussian mode would present special spiral fringes so that the topological charges of these OAM modes can be revealed. The number and rotational direction of the fringes are corresponding to the value and symbol (±) of $l$, respectively[4]. Fig.S2 shows the interference patterns for azimuthally polarized OAM modes with $l=2\sim6$ and $l=-6\sim-1$. A tunable laser is adopted as the input of the OAM emitter and the corresponding wavelengths are marked in the bottom. Each of OAM modes could be decomposed into left-handed circular polarized (LHCP) component with $l_L=l+1$ and right-handed circular polarized (RHCP) component with $l_R=l-1$. The topological charges of the azimuthally polarized OAM modes could be confirmed by both interference patterns.

This method could also utilized to certify the topological charges of OAM modes switched by varied driving voltage. In Fig.S3, when incident wavelength is fixed at signal wavelength (1557.36nm/1547.72nm), the generated OAM modes as different driving voltages are measured. The results are similar to those in Fig. S2 but the tuning mechanism of $l$ is different.

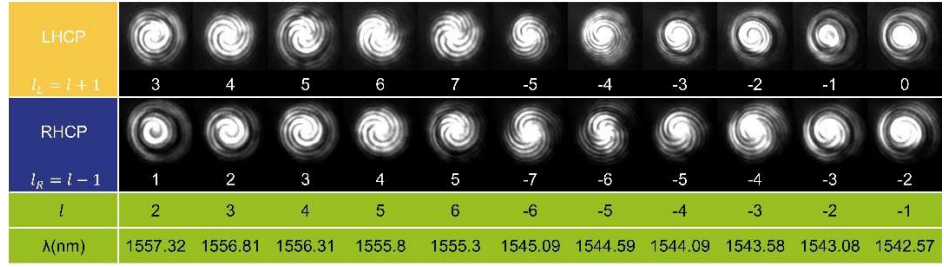

Fig. S2. The interference patterns of 11 OAM modes and the fundamental Gaussian mode. Each OAM mode corresponds two interference patterns with topological charges ($l_L$ and $l_R$) marked below them. The last column shows the operating wavelengths of the corresponding OAM modes.

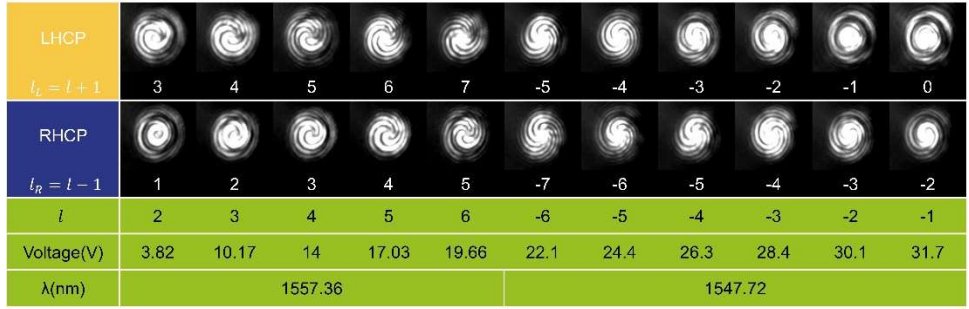

Fig. S3. The interference patterns of OAM modes whose topological charges are switched by the driving voltages. The incident wavelength is fixed at 1557.36nm for $l$=2~6 and 1547.72nm for $l$=-6~-1.

## 4. The arrangement of signal and idler wavelength

The photon pairs are generated through the degenerate spontaneous four-wave mixing (SFWM) in SW-WG. In this work, the pump wavelength is set as 1552.5nm according to our experimental conditions. As shown in Fig. S4 (a), the group velocity dispersion (GVD) of SW-WG has been simulated. The pump wavelength ($\lambda_p$) of 1552.5nm is marked as the purple dotted line. Although the GVD at $\lambda_p$ is not zero, the length of SW-WG is only 5mm and the total GVD is $2.355\times10^{-3}$ps/nm. Thus, the phase matching condition of SFWM is approximately satisfied[5,6]. When the driving voltage is 0V, $\lambda_p$ is near the wavelength (1552.71nm) corresponding to $l$=11 of OAM emitter. Furthermore, the signal and idler wavelengths ($\lambda_s$, $\lambda_i$) are distributed equally on both sides of the $\lambda_p$. Here, $\lambda_s$ and $\lambda_i$ should be not too far from $\lambda_p$ to guarantee enough generation rate of photon pairs[6] and as close as possible to wavelengths corresponding to $l=\pm1$ ($\lambda_{l=1}$=1557.83nm, $\lambda_{l=-1}$=1542.57nm), which would be helpful to reduce the driving voltages for the alignments of $\lambda_s$ and $\lambda_{l=\pm1}$. Furthermore, signal and idler photons would pass through a standard DWDM before detection so that the Raman noise can be filtered out. Thus, the $\lambda_s$ and $\lambda_i$ have to be aligned to the operating wavelength of the DWDM. Actually, 1558.17nm and 1545.92nm are the first choice, since they can meet both two requirements mentioned above. However, when the operating wavelength of $l$=2 is aligned with $\lambda_s$=1558.17nm through the thermo-optical controller, it can be seen from the Fig. S4 (b) that the pump and idler photons are partially coupled into the OAM emitter, which would decrease the CCs and CAR. Hence, $\lambda_s$ and $\lambda_i$ are finally set as 1557.36nm and 1547.72nm while the cost is the absence of the OAM mode with $l$=1.

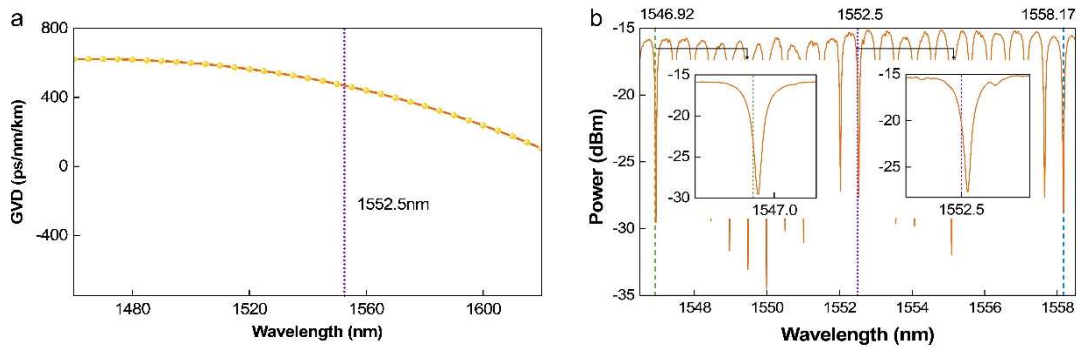

Fig.S4 a. The simulation result of the SW-WG's group velocity dispersion (GVD) versus wavelengths. b. The relation between pump/signal/idler and the transmission spectrum of OAM emitter when the operating wavelength of l=2 is aligned with λs=1558.17nm.

## 5. Measurement of mode purity

The mode purity presents the power ratio of the target OAM mode among actual generated modes. Obviously, the mode purity of a target OAM mode should be as near unity as possible. As shown in the Fig.S5 (a), the measurement can be regarded as the tomography for target OAM mode and the spatial light modulator (SLM) serves as the mode filter. Due to the special helical phase structures, light beams with OAM possess a doughnut-like intensity cross-sections (vertical to the propagation direction)[4] and could not be coupled into the single mode fiber (SMF). Here, only the OAM mode with $l=-l_{mask}$ corresponding to topological charge of the settled phase mask on SLM ($l_{mask}$) could be converted to fundamental Gaussian mode and detected by the SMF-packaged power meter. However, the SLM operates with linear polarization incidence but the generated OAM modes are azimuthally polarized. Thus, a quarter wave plate (QWP) and a polarizer are utilized to transform OAM modes from azimuthal polarization to linear polarization.

Specifically, the incident wavelength is fixed at the signal wavelength (1547.72nm or 1557.36nm) and different OAM modes could be obtained by properly setting the driving voltage. For a target OAM mode, a SLM with phase masks of $l_{mask}$=-7~7 is utilized to extract the corresponding modes and 15 values of optical power could be obtained. The mode purity is calculated as the power ratio of target OAM mode over the sum of all 15 values. It should be noticed that one component of the LHCP with $l_L=l+1$ or RHCP with $l_R=l-1$ is firstly extracted from the azimuthally polarized mode with topological charge of $l$ and then transformed to the linear polarization before SLM. Thus, the relation of the topological charge of the phase mask on SLM versus the OAM mode is $l_{mask}=-l_L=-(l+1)$ for LHCP component or $l_{mask}=-l_R=-(l-1)$ for RHCP one. Fig.S5 (b) shows a concrete example, in which the horizontal axis is the $l_{mask}$ and the corresponding RHCP component of the OAM mode is extracted. With the relation of $l_{mask}=-l_R=-(l-1)$, the actual topological charge of the corresponding OAM mode should be $l=-l_{mask}+1$. The transverse intensity profiles after the SLM are shown in Fig.S5 (c) with the corresponding $l_{mask}$ marked below. It can be seen that the main OAM mode is $l_R=-2$ ($l=-1$) and the mode purity is 82.62%. Here, the measured power is normalized by the total power of all OAM modes. For 11 OAM modes, there are two sets of mode purities as shown in Fig. S5 (d) corresponding to LHCP and RHCP components. The results shown in Fig.4 of the main context are the average values of both LHCP and RHCP components.

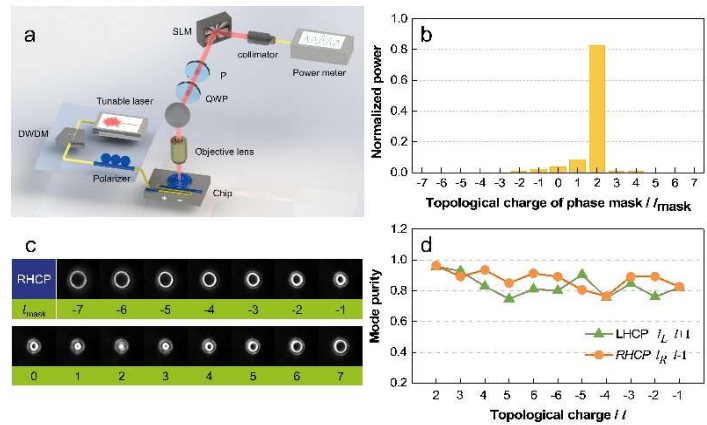

Fig.S5 a. The experimental scheme for measuring the mode purities of generated OAM modes. b. A typical result of OAM mode with $l_R$=-2. There are 15 values of moralized power and the value corresponding to $l_{mask}$=2 is the required mode purity. c. The transverse intensity profiles of OAM mode after different masks which are marked below. d. The measured mode purities of both LHCP and RHCP components corresponding to 11 OAM modes.

## 6. The insertion loss of detection system

The insertion loss introduced by the detection system includes the transmission loss of the objective lens, wave plates (a quarter wave plate and a polarizer), spatial light modulator (SLM) and collimator. Among them, the insertion loss of the objective lens is constant as -3.98dB. Other losses are measured for $l$=2~6 and -6~-1 under the fixed incident wavelength of 1557.32nm and 1547.72nm, respectively. The results are shown in Fig.S6 (a) and each value is averaged by two linear polarization (the azimuthal polarization is converted to horizontal and vertical polarization). For 11 OAM modes, the losses of wave plates and SLM are around ~-4dB and nearly independent on the topological charges. For the insertion loss of collimator, it is obtained by subtracting the received power after collimator from total power before collimator. Since only the fundamental Gaussian mode can pass through the collimator, thus, the measured values shown in Fig.S6 (a) actually contain the effect due to different mode purities of different OAM modes. By comparing the insertion loss of collimator (green dots) in Fig.S6 (a) and the value of mode purities shown in Fig.4 of the main context, it can be found that both curves have the similar variation trend. Moreover, after deducting the insertion losses of detection system, the coincidence counts (CCs) are calculated and shown in the Fig.S6 (b)

(Fig.7 in main context). This results could present the actual generation rate of the heralded single photons with our device.

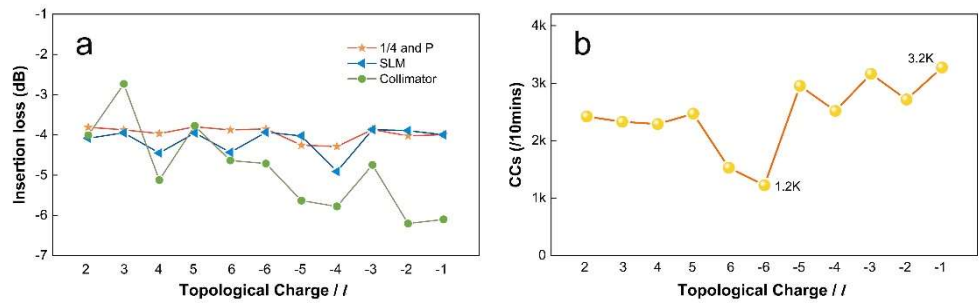

Fig.S6 a. The insertion losses of wave plates, spatial light modulator (SLM) and collimator for 11 OAM modes. b. The estimated coincidence counts (CCs) of heralded single photons after deducting the insertion loss of detection system.